%% file: body.tex
\documentclass[11pt,twocolumn]{article}

\usepackage{amsmath}
\usepackage{amssymb}
\usepackage{parskip}
\usepackage{pgfplots}
\usepackage{multirow}
\usepackage{tikz}
\usepackage{hyperref}
\usepackage{flushend} % references two column leveling
\usepackage{authblk}
\usepackage{algorithm}

\usepackage[margin=0.75in]{geometry}

\usepackage{color}
\definecolor{lightgray}{rgb}{.9,.9,.9}
\definecolor{darkgray}{rgb}{.4,.4,.4}
\definecolor{purple}{rgb}{0.65, 0.12, 0.82}

\usepackage{listings}
\lstdefinelanguage{javascript}{
  basicstyle=\ttfamily\tiny,
  breaklines=false,
  tabsize=2,
  keywords={int, float, function, return, var, let, const, for, while, if, break, true},
  keywordstyle=\color{blue}\bfseries,
  ndkeywords={class, export, boolean, throw, implements, import, this},
  ndkeywordstyle=\color{darkgray}\bfseries,
  identifierstyle=\color{black},
  sensitive=false,
  comment=[l]{//},
  morecomment=[s]{/*}{*/},
  commentstyle=\color{purple}\ttfamily,
  stringstyle=\color{red}\ttfamily,
  morestring=[b]',
  morestring=[b]"
}

\title{Theoretical and Empirical Analysis of a Fast Algorithm for Extracting Polygons from Signed Distance Bounds}
\author{Nenad Marku\v{s}\thanks{nenad.markus@fer.hr}}
\author{Mirko Su\v{z}njevi\'{c}\thanks{mirko.suznjevic@fer.hr}}
\affil{University of Zagreb Faculty of Electrical Engineering and Computing\\Unska 3, 10000 Zagreb, Croatia}

\date{}

\begin{document}
	\maketitle

	\begin{abstract}
        Recently there has been renewed interest in signed distance bound representations due to their unique properties for 3D shape modelling.
        This is especially the case for deep learning-based bounds.
        However, it is beneficial to work with polygons in most computer-graphics applications.
		Thus, in this paper we introduce and investigate an asymptotically fast method for transforming signed distance bounds into polygon meshes.
		This is achieved by combining the principles of sphere tracing (or ray marching) with traditional polygonization techniques, such as Marching Cubes.
		We provide theoretical and experimental evidence that this approach is of the $O(N^2\log N)$ computational complexity for a polygonization grid with $N^3$ cells.
		The algorithm is tested on both a set of primitive shapes as well as signed distance bounds generated from point clouds by machine learning (and represented as neural networks).
		Given its speed, implementation simplicity and portability, we argue that it could prove useful during the modelling stage as well as in shape compression for storage.

		The code is available here: \url{https://github.com/nenadmarkus/gridhopping}.
	\end{abstract}


	\section{Introduction}
  		A Signed Distance Field (SDF) for some shape $S$ is a function $f_S:\mathbb{R}^3\rightarrow\mathbb{R}$ such that $f_S(x, y, z)$ returns a geometric distance from the point $(x, y, z)\in\mathbb{R}^3$ to $S$.
		If $(x, y, z)$ is within $S$, then the returned distance is negative.
		In practice we often cannot obtain the exact distance to the shape, but work with distance bounds that underestimate the distance some of the time and never overestimate it.
		This enables us to efficiently represent a larger class of shapes.
        J. C. Hart \cite{Hart94spheretracing} (see also \cite{HartEtAl89rtfractals}) uses the definition that $f_S$ is a signed distance bound (SDB) of $S$ if and only if for all $\mathbf{x}\in\mathbb{R}^3$ we have
		$$
			\vert f_S(\mathbf{x})\vert\leq
			\min_{\mathbf{y}\in f_S^{-1}(0)}\vert\vert\mathbf{x} - \mathbf{y}\vert\vert_2
		$$
		where $f_S^{-1}(0)=\{\mathbf{z}: f_S(\mathbf{z})=0\}$.
		The earlier mentioned sign rule applies to the interior of $S$.
        It should be noted that the term SDF is often used in the literature even though SDB is more appropriate \cite{Hart94spheretracing,HartEtAl89rtfractals}.
  
        Signed distance bound representations show significant potential due to their rendering speed, storage efficiency, implementation simplicity and shape distribution modelling capabilities.
        They are powerful tools which are used in computer graphics, simulation, and medical imaging domains.
  		Here are some typical applications:
		\begin{itemize}
			\item
				shape compression --- it may be more storage-efficient to store a signed distance bound (e.g., represented by a small neural network \cite{davies2020overfit}) than as a triangle mesh;
			\item
				statistical shape representations for generative modelling --- sampling from such a representation \cite{GenShapeCVPR2019,DeepShapeCVPR2019} can potentially help add variety to virtual worlds;
            \item
                shape recovery from point clouds and noisy measurements \cite{ShapeFromPtCould,SecretsOfWildSDFs2021};
			\item
				shape manipulation --- such as in the paper by Hao et al. \cite{hao2020dualsdf};
			\item
				computer-aided design (CAD).
		\end{itemize}
  		
        In most graphics applications (e.g., classical 3D game engines), due to GPU hardware specifics and well-developed prior practices, it is beneficial to work with polygons even though some assets are encoded as SDBs.  
        Therefore, converting a signed distance bound into a polygon mesh warrants investigation\footnote{The inverse process has been studied more extensively, e.g., \cite{DoubleLayerSDF,GenShapeCVPR2019}}.

        The objective of this paper is to present and evaluate a method for transforming signed distance bounds into polygon meshes.
        The contributions are:
        \begin{enumerate}
            \item we describe a fast algorithm for polygonizing SDBs based on sphere tracing (ray marching),
            \item we theoretically show that this algorithm is asymptotically faster than the obvious approaches (e.g., Marching cubes \cite{LorensenCline87marchingcubes} over the whole grid),
            \item we empirically confirm this in practice for signed distance bounds encoded with deep neural networks.
        \end{enumerate}

		The next section provides context, basic definitions and related prior work. The description of our method and its analysis/comparisons follows after that.

	\section{Basic definitions and related work}\label{sec:overview}
        In our case, the polygonization volume is an axis-aligned cube centered at the origin.
		The shape represented by its SDB is withing this cube.
		This cube is partitioned into a rectangular grid with $N^3$ cells by subividing each of its sides into $N$ intervals of equal size.
		If we assume we are not dealing with fractals, only $O(N^2)$ cells asymptotically contain the surface of our shape.
		Thus, the only triangles that need to be computed are passing through these cells.
		This puts the lower bound on the complexity of the triangulation algorithm.
		However, the challenge is to isolate just these $O(N^2)$ cells.

		The simplest solution, which leads to $O(N^3)$ complexity, is to check each of the $N^3$ cells.
		This process is called \textbf{enumeration} \cite{bloomenthal} (Section 4.2.2 in the book) and for some purposes it is too slow.
		The original applications of the Marching cubes algorithm \cite{LorensenCline87marchingcubes} applied this slow approach.
		However, these applications usually dealt with real-world data obtained through measurments (e.g., CT scans), not well defined mathematical objects.

		An improvement to this basic enumeration scheme is known as \textbf{continuation} \cite{bloomenthal} (Section 4.2.3 in the book).
		This idea of this approach is quite simple.
		Starting from a single "seed" cell that intersects the surface of the shape, new cells are propagated across the surface until the entire surface is enclosed.
		The complexity of this algorithm is $O(N^2)$ because only the surface cells are analyzed.
		Thus, we also call it surface tracing/crawling.
		Its main disadvantage is that it produces a single mesh component for each seed cell.
		I.e., we need a separate seed cell to polygonize all disjoint shape components.
		This can be problematic in practice.
		Another issue with this approach is caused by limited numerical precision of digital computers when representing seed cell parameters
		(e.g., their center point coordinates) -- this complicates the positioning of the seed cell.
		Of course, this problems only show up for high grid resolutions, i.e., large values of $N$.

		We also mention the unpublished work of Christopher Olah \cite{olah2011manipulation,olahImplicitCAD}. This method aims at similar goals (transform an SDB into a triangle mesh), but uses a different subdivision strategy over the grid (recursive partitioning instead of sphere tracing) and no detailed complexity analysis was performed.
        It is not a formal academic work, but an implementation.

        Besides the mentioned work that warrants a direct experimental comparison (enumeration and continuation \cite{bloomenthal}, Marching cubes \cite{LorensenCline87marchingcubes}) to our method, we also mention more broad work for additional context. There are a number of algorithms for speeding up isosurface extraction (as polygon meshes) from real-world data such as MRI and CT scans. E.g., based on octrees \cite{WilhelmsGelder} or other more exotic data structures \cite{CigoniEtAl,LivnatShenJohnson}.
        However, these methods assume that data points on the grid are already known (sampled).
        Thus, direct application of these algorithms to polygonizing SDBs is not possible.

        There is also a large body of work of generative models for geometry.
        In this context, we mention those that directly output polygons. Gao et al. \cite{gao2022get3d} represent geometry as a signed distance field defined on a deformable tetrahedral grid and incorporate this into a learning framework.
        Their end result is a neural representation that can directly generate polygonal models from the training data distribution.
        In the PolyDiff paper \cite{alliegro2023polydiff}, the authors introduce a mesh generator based on a denoising diffusion process \cite{sohldickstein2015deep,ho2020denoising}.
        In the MeshGPT paper \cite{siddiqui2023meshgpt}, the authors use a decoder-only transformer neural network to directly output the polygons of a mesh.
        All these works are learning-based, i.e., require a large dataset in order to be effective, are not easy to implement/use in a production system like a game engine and require significant computational resources.

		The next section provides a detailed (mathematical, algorithmic) description of the gridhopping method to enable its analysis.

	\section{Method}\label{sec:ghop}
		Another way to speed up the triangulation process is to apply a variant of ray marching called Sphere Tracing (described by John C. Hart \cite{HartEtAl89rtfractals,Hart94spheretracing}).
		Note that sphere tracing is originally a 3D rendering scheme.
		However, we re-purpose it here for polygonization.
		The basic idea is to define a ray and move along its direction until you find the intersection with the shape or exit the rendering volume.
		In sphere tracing, the marching step is set to be equal to the (estimated) distance of the current point to the shape.
		This approach greatly speeds up the process of finding the intersection.
		However, unlike in the ray marching-based rendering of images, we do not stop the marching process at the surface.
		The marching along the ray in continued (starting at the next cell along direction of the ray) until the end of the polygonization volume is reached.
		Let us call this method \textbf{gridhopping}.
		This process enables us to efficiently isolate the $O(N^2)$ cells that contain the surface of the shape:
		we present theoretical and experimental evidence that this method extracts a triangle mesh from a signed distance bound in $O(N^2\log N)$ steps for a grid with $N^3$ cells.
		Note that this is a significant speed improvement over the basic approach that analyses each of the $N^3$ cells, leading to $O(N^3)$ complexity.

		Without loss of generality, we assume that our polygonization volume is a unit cube centered at the origin.
		The grid resolution is specified by $N$: there are $N^3$ cubic cells in the grid, each with a volume equal to $\frac{1}{N^3}$.
		Each cell is assigned a triplet of integers $(i, j, k)$ with $i, j, k\in \{0, 1, 2, \ldots, N-1\}$.
		The centroids of the cells are computed according to the following rules:
		\begin{align}\label{eq:cells}
			x_i &= -\frac{1}{2} + \frac{1}{2N} + \frac{i}{N}\nonumber\\
			y_j &= -\frac{1}{2} + \frac{1}{2N} + \frac{j}{N}\\
			z_k &= -\frac{1}{2} + \frac{1}{2N} + \frac{k}{N}\nonumber
		\end{align}

		A total of $N^2$ rays are cast in the $+z$ direction from the plane $z=-0.5+\frac{1}{2N}$.
		Such rays have the following vector parameterization for $\lambda \geq 0$:
		\begin{equation}
			R_{ij}\;\;\ldots\;\;\mathbf{r}=
			\mathbf{o}_{ij} + \lambda\mathbf{d}
		\end{equation}
		with $\mathbf{o}_{ij}=(x_i, y_j, -0.5+\frac{1}{2N})^T$ is the origin of ray $R_{ij}$ and $\mathbf{d}=(0, 0, 1)^T$ is its direction.
		The $(x_i, y_j)$ pairs ($N^2$ of them) are computed according to equations \eqref{eq:cells}.

		We move along each ray using the ray marching (sphere tracing) \cite{HartEtAl89rtfractals,Hart94spheretracing} method.
		If the polygonization volume contains a shape $S$ described by its signed distance bound $f_S$, the following iteration describes this process:
		\begin{equation}\label{eq:iter}
			\mathbf{r}_{n+1}=
			\mathbf{r}_n + \left|f_S(\mathbf{r}_n)\right|\mathbf{d}
		\end{equation}
		The iteration starts at $\mathbf{r}_0=\mathbf{o}_{ij}$ and continues until $\left|f_S(\mathbf{r}_n)\right|$ is sufficiently small
		(indicating we are very close to the surface of $S$, by definition of $f_S$).
		In our case, we are only interested to move close enough to the surface to determine the $(i, j, k)$ triplet determining the cell.
		Simple algebra shows that a cell possibly intersects the surface of $S$ and we have to call a polygonization routine if the distance $\left|f_S\right|$ is less than or equal to
		\begin{equation}
			\sqrt{
				\left(\frac{1}{2N}\right)^2 + \left(\frac{1}{2N}\right)^2 + \left(\frac{1}{N}\right)^2
			}=
			\frac{\sqrt{6}}{2N}
            \label{eq:polydist}
		\end{equation}
		The pseudocode in \ref{code:method} contains the details, and Figure \ref{visualsummary} provides a visual summary/illustration.
        \begin{algorithm}
			\label{code:method}
			\resizebox{0.52\textwidth}{!}
			{
\begin{lstlisting}[language=JavaScript]
// inputs:
//  * `eval_sdb` is the signed distance bound represeting a shape
//  * `N` is the grid resolution
function apply_grid_hopping(eval_sdb, N)
{
	for (int i=0; i<N; ++i)
		for (int j=0; j<N; ++j)
		{
			int k=0;
			while (true)
			{
				// set the origin of the ray
				float x=-1.0/2.0+1.0/(2.0*N)+i/N;
				float y=-1.0/2.0+1.0/(2.0*N)+j/N;
				float z=-1.0/2.0+1.0/(2.0*N)+k/N;
				// use ray marching to determine how much to move along the ray
				float t = trace_ray(
					[x, y, z],             // origin of the ray
					[0.0, 0.0, 1.0],       // direction of the ray
					eval_sdb,              // signed distance bound
					1.05*(1.0/2.0 - z),    // max distance to travel
					Math.sqrt(6.0)/(2.0*N) // distance to surface we require
				);
				// set the new value of z and its associated cell, (i, j, k)
				z = z + t;
				k = Math.floor(N*(z + 1.0/2.0  - 1.0/(2.0*N)));
				// are we outside the polygonization volume?
				if (k>N-1 || z>1.05/2.0)
					break;
				// polygonize cell (i, j, k)
				... // <- polygonizaiton code goes here, e.g., Marching cubes
				// move further along the z direction
				++k;
			}
		}
}
\end{lstlisting}
			}
            \caption{Pseudocode for gridhopping.}
		\end{algorithm}

		If the ray intersects the surface and we denote the closest intersection to $\mathbf{r}_0$ with $\mathbf{r}^*$,
		then the above iteration converges to $\mathbf{r}^*$.
		This is because
		\begin{enumerate}
			\item
				$\left|f_S(\mathbf{r}_n)\right|\geq 0$;
			\item
				 on the ray between $\mathbf{r}_0$ and $\mathbf{r}^*$, $f_S(\mathbf{r})=0$ only for $\mathbf{r}=\mathbf{r}^*$;
			\item
				the iteration will never "overshoot" $\mathbf{r}^*$ because $f_S$ is a signed distance bound.
		\end{enumerate}
		See \cite{Hart94spheretracing} for additional analysis.

        \begin{figure*}
        \center
        \resizebox{0.7\textwidth}{!}
		{
        \begin{tikzpicture}[line cap = round]
            \clip (-12, -3.5) rectangle (0.5, 4.5);
            % shape filled with red
            \fill[red] plot[smooth cycle, tension=.5] coordinates {(1.0, -2.0) (-1.0, 0.0) (-5.0, -2.0) (-6.0, 0.0) (2.0, 4.0) (5.0, -1.0) (4.0, -3.0) (2.0, -1.0)};
            % grid with thicker and dashed lines
            \draw[step=1.0,gray!25!,very thick, dashed] (-12, -4) grid (6, 5);
            % z axis
            \draw[->, line width=1.0mm] (-11.5, 0.5) -- (-7.5, 0.5); % z-axis arrow
            \node[below,font=\Huge] at (-8.5, 0.25) {$z$}; % z-axis label
            % xy plane
            \draw[-, line width=1.0mm] (-9.5, -2.5) -- (-9.5, 3.5); % xy-plane arrow
            \node[above,font=\Huge] at (-9.5, 3.5) {$xy$}; % xy-plane label
            % cell that gets polygonized first
            \fill[blue!20, opacity=0.5] (-5,1) rectangle ++(1,1);
            % ray
            \draw[->, dashed, green!70!black, line width=1.0mm] (-15.5, 1.5) -- (-1.5, 1.5);
            % circles
            \draw[green!70!black, dashed, line width=0.4mm] (-10.5, 1.5) circle [radius=4.6];
            \draw[green!70!black, dashed, line width=0.4mm] (-5.9, 1.5) circle [radius=1.15];
            % locations along the ray
            \fill[green!70!black] (-10.5, 1.5) circle [radius=0.15];
            \fill[green!70!black] (-5.9, 1.5) circle [radius=0.15];
            \fill[green!70!black] (-4.75, 1.5) circle [radius=0.15];
        \end{tikzpicture}
        }
        \caption
        {
            A 2D illustration of the gridhopping method: grid cells are dashed gray squares, the shape with a SDF is in red, the $z$ axis and the $xy$ plane are in black.
            The dashed green line (parallel to the $z$ axis) represents one of the rays along which sphere tracing is applied.
            Since the SDF of the red shape is known, we can compute the maximum step size along the ray: the green dots are locations obtained from these calculations.
            The Marching cubes polygonization is first applied to the light blue cell since it is the first one along the ray that satisfies the condition related to equation \ref{eq:polydist}.
        }
        \label{visualsummary}
        \end{figure*}

	\section{Theoretical analysis of computational complexity}\label{sec:complexity}
		We analyze the asymptotic number of steps required by the method from previous section to polygonize a shape defined through its signed distance bound.
		For non-fractal shapes, there are at most $O(N^2)$ cells that contain polygons.
		The challenge is to isolate these cells in a fast manner.
		The trivial way is to check all $N^3$ cells.
		This may be too slow for some applications when high resolution (large $N$) is required.
		Our claim is that the algorithm from the previous section is faster than that:
		its complexity is $O(N^2\log N)$.

		We provide evidence for this in the following steps:
		\begin{enumerate}
			\item
				provide a proof for polygonizing planes;
			\item
				provide a proof for polygonizing axis-aligned boxes;
			\item
				argue that any non-fractal shape can be approximated as a union of boxes.
		\end{enumerate}
		These steps are explained in the following three subsections.

		\subsection{Polygonizing planes}
			A plane is a flat, two-dimensional surface that extends infinitely far.
			Of course, we are interested in polygonizing only the part that intersects with the polygonization volume.

			The exact signed distance from a point $\mathbf{r}$ to a plane $P$ is given by the following equation \cite{ppdist}:
			\begin{equation}\label{eq:plane}
				D_P(\mathbf{r})=
				\mathbf{n}_P^T\cdot (\mathbf{r} - \mathbf{r}_P)
				,
			\end{equation}
			where $\mathbf{r}_P$ is some point lying on $P$ and $\mathbf{n}_P$ is $P$'s normal vector such that $\mathbf{n}_P^T\cdot\mathbf{n}_P=||\mathbf{n}_P||_2^2=1$.

			We analyze three different cases: two cases of axis-aligned planes and one case for a plane in general position.
			The first case is when the plane and the rays are perpendicular.
			In our case, since the rays are cast in the $+z$ direction, this corresponds to the plane $z=C$ for some constant $C$.
			The second case is when the plane and the rays are parallel (plane specified by $x=C$ or $y=C$).
			The third case is the plane in a general position.

			\textbf{Case \#1: plane $P$ and rays are perpendicular}.
			First, notice that the orientation of the plane does not matter since the ray marching always uses the absolute value of the computed distance bound.
			Thus, we have two sub-cases: approaching the plane and escaping the vicinity of its surface.
			The approaching phase is performed in a single step for each ray since $D_P$ provides the exact distance estimate.
			Hence, its complexity is $O(1)$, independent of $N$, the position of the plane along the $z$ axis and the starting point.
			Since there are $N^2$ rays, the complexity of the approaching phase is $O(N^2)$.
			Escaping the plane's surface requires more work and the following analysis holds for each of the $N^2$ rays that need to be cast.
			Let $\mathbf{r}_0$ denote the starting point in the vicinity of $P$'s surface.
			Note that $D_P(\mathbf{r}_0)$ is about $\frac{1}{N}$ in size immediately after the polygonization routines for $P$'s cells have been invoked
			(at most two in this case, only one containing polygons).
			Analyzing the iteration \eqref{eq:iter}, it is easy to see that $D_P(\mathbf{r}_1)=2\cdot D_P(\mathbf{r}_0)$ and in general the following holds:
			\begin{equation}
				D_P(\mathbf{r}_n)=2^n\cdot D_P(\mathbf{r}_0)
				,
			\end{equation}
			i.e., the method escapes the surface in steps of exponentially increasing size.
			If $D_P(\mathbf{r}_0)$ is about $\frac{1}{N}$, then the number of steps required to exit the polygonization volume is $O(\log N)$.
			Given that there are $N^2$ such rays, the complexity of the escaping phase is $O(N^2\log N)$.
			The approaching and escaping phase are performed sequentially.
			Thus, the complexity of polygonizing a plane in this scenario is $O(N^2\log N)$.

			\textbf{Case \#2: plane $P$ and rays are parallel}.
			First, notice that if the distance between a ray and $P$ is equal to $\frac{k}{N}$ in this case, then the method exits the polygonization volume in approximately $\frac{N}{k}$ steps.
			There are $N$ rays marching through cells that contain $P$.
			The algorithm takes approximately $N$ steps along each of these rays before terminating.
			Next, notice that there are $N$ rays above and $N$ rays below $P$, parallel to $P$ and of distance approximately $\frac{k}{N}$ to $P$ for some integer $k < N$.
			These rays require about $N/k$ steps before exiting the polygonization volume.
			Thus, a conservative estimate for the number of steps $S$ for all $N^2$ rays is
			\begin{multline}
				S\leq
				N^2 + 2\left( N^2 + \frac{N^2}{2} + \frac{N^2}{3} + \cdots + \frac{N^2}{N-1}  + N \right)\\
				=N^2\cdot\left(1 + 2H_N\right)
				,
			\end{multline}
			where $H_N$ is $N$th partial sum of the harmonic series \cite{hseries}: $H_N=\sum_{n=1}^N \frac{1}{n}$.
			The number $H_N$ is about as large as $\log N$.
			The reason for this comes from the comparison of $H_N$ and the integral $\int_1^{N}\frac{1}{x}\mathop{dx}$,
			which can be solved analytically.
			Thus it follows that $S\in O(N^2\log N)$ and the complexity of case \#2 is $O(N^2\log N)$.

			\textbf{Case \#3: the general case}.
			Due to easier exposition and without loss of generality, we assume that the plane passes through origin (i.e., $\mathbf{r}_P=\mathbf{0}$).
			Combining this assumption with equations \eqref{eq:iter} and \eqref{eq:plane}, we get the following iteration for the $z$ coordinate:
			\begin{equation}
				z_{n+1}=
				z_n + \left| n_x x_0 + n_y y_0 + n_z z_n \right|
			\end{equation}
			where $(n_x, n_y, n_z)^T$ is the unit normal of the plane and $(x_0, y_0, z_0)^T$ is the origin of the ray.
			Let us denote with $z^*$ the intersection of the ray and the plane:
			\begin{equation}
				z^*=
				-\frac{n_x x_0 + n_y y_0}{n_z}
			\end{equation}
			Without loss of generality, we assume that $n_z<0$.
			There are two sub-cases:
			(1) the method approaches the plane along the ray and (2) the method moves away from the plane along the ray.
			In the first sub-case, we have $n_x x_0 + n_y y_0 + n_z z_n > 0$.
			In this scenario, it is easy to see that
			\begin{equation}
				z^* - z_n =
				(1 + n_z)\cdot (z^* - z_{n-1})=
				\cdots=
				(1 + n_z)^n (z^* - z_0)
			\end{equation}
			Since $1 + n_z$ is between $0$ and $1$, we have that the number of iterations $n$ has to be about $O(\log N)$ so that $D_P(\mathbf{r}_n)$ becomes less than $\frac{\sqrt{6}}{2N}$
			(at which point the polygonization routine is invoked and we can move to the other side of the plane).
			In the second sub-case, we have $n_x x_0 + n_y y_0 + n_z z_n < 0$.
			Now the following holds:
			\begin{equation}
				z_n - z^* =
				(1 - n_z)\cdot (z_{n-1} - z^*)=
				\cdots=
				(1 - n_z)^n (z_0 - z^*)
			\end{equation}
			Since $1 - n_z$ is greater than $1$, at most $O(\log N)$ iterations along the ray are needed to exit the polygonization volume.
			Given that there are $N^2$ rays in total, the complexity of case \#3 is $O(N^2\log N)$.

		\subsection{Polygonizing rectangular boxes}
			A rectangular box can be obtained by intersecting six axis-aligned planes.
			Let $d_1, d_2, \ldots, d_6$ be the distances from point $(x, y, z)$ to each of these planes.
			Then the distance to the box is bounded by
			\begin{equation}\label{eq:boxdist}
				f(x, y, z)=
				\max\{d_1, d_2, d_3, d_4, d_5, d_6\}
			\end{equation}
			Since polygonizing each of the box sides takes $O(N^2\log N)$ steps, this is also the total complexity of polygonizing a box.

			This can also be justified by the fact that the $\max$ operation partitions the polygonization volume into several regions.
			In each of these regions only the distance to one particular plane is relevant (largest $d_i$).
			All the rays passing through this region require at most $O(N^2\log N)$ steps before exiting the region.
			The conclusion about the total complexity follows from the fact that the number of such regions is finite.

			As noted, Equation \eqref{eq:boxdist} bounds the distance to the box.
			The exact distance function can be constructed and this leads to more efficient marching in practice
			(a constant speed-up, not in the asymptotic sense)
			\footnote{For example, see \url{https://www.youtube.com/watch?v=62-pRVZuS5c}.}.

		\subsection{Polygonizing other shapes}
\begin{figure}
				\centering
				\resizebox{0.5\textwidth}{!}
				{
					\input{boxapprox.tikz}
				}
				\caption
				{
					One possible approximation of a shape (red) as a union of axis-aligned boxes (blue).
					Note that the approximation would be much more efficient if non-axis-aligned planes are used on the boundary.
				}
				\label{fig:boxapprox}
			\end{figure}
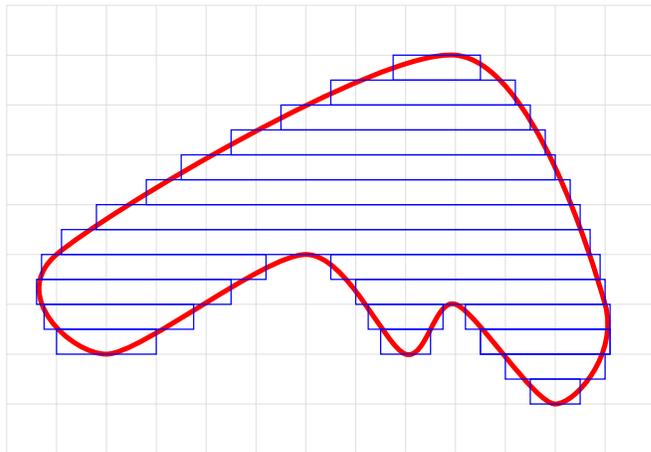

			A shape can be approximated by $K$ axis-aligned boxes.
			See Figure \ref{fig:boxapprox} for an illustration of this process.
			Of course, we can improve the quality of approximation by increasing $K$.
			It is important to note that $K$ does not depend on grid resolution $N$.
			The efficiency of approximation can be increased by using non-axis-aligned planes at the boundary of the shape.
			This process is not unlike the use of triangle meshes in modern computer graphics.

			Approximating the shape as a union of $K$ boxes keeps the $O(N^2\log N)$ polygonization complexity.
			This is because the union of $K$ boxes (and, in general, shapes) can be obtained by applying the $\min$ operation to combine all the individual distance bounds.
			The number of steps the method has to make in this case is asymptotically no worse than polygonizing each box on its own.
			Thus, the total number of steps scales as $O(N^2\log N)$ since the grid resolution $N$ does not depend on $K$.

	\section{Experimental analysis}\label{sec:experiments}
		In this section we experimentally compare the three methods\footnote{Refer to Section \ref{sec:overview} for the overview of the methods.}: gridhopping (\texttt{ghop}), enumeration (\texttt{enum}) and continuation (\texttt{cont}).
		The goal is to show that the \texttt{enum} method is asymptotically slower than \texttt{cont} and \texttt{ghop}.

		The polygonization of a cell is obtained with the Marching cubes algorithm.
		To achieve this, we implement all methods and run them on different scenes for varying grid resolutions.
		It is important to note that all implementations produce exactly the same meshes when polygonizing signed distance bounds.

		\subsection{Primitives and simple shapes}\label{sec:experiments-primitive}
\begin{figure}
			\centering
			\resizebox{0.5\textwidth}{!}
			{
				\includegraphics{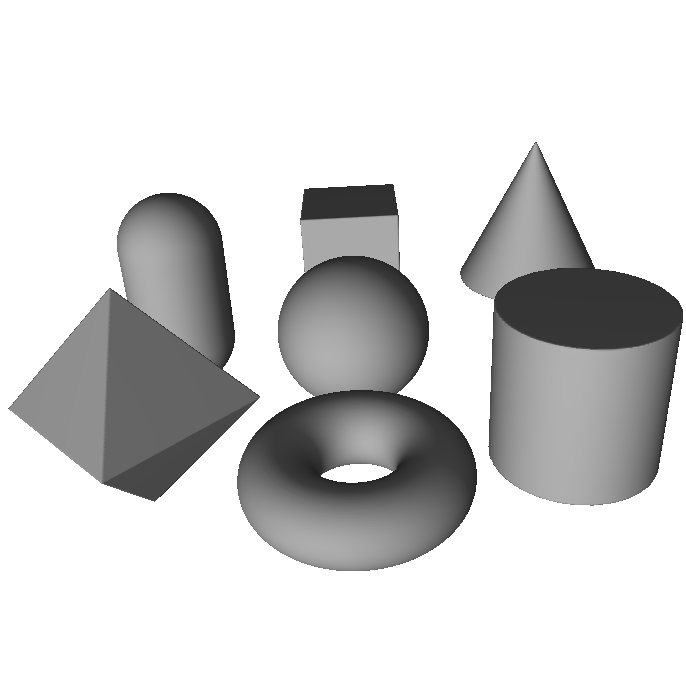}
				\includegraphics{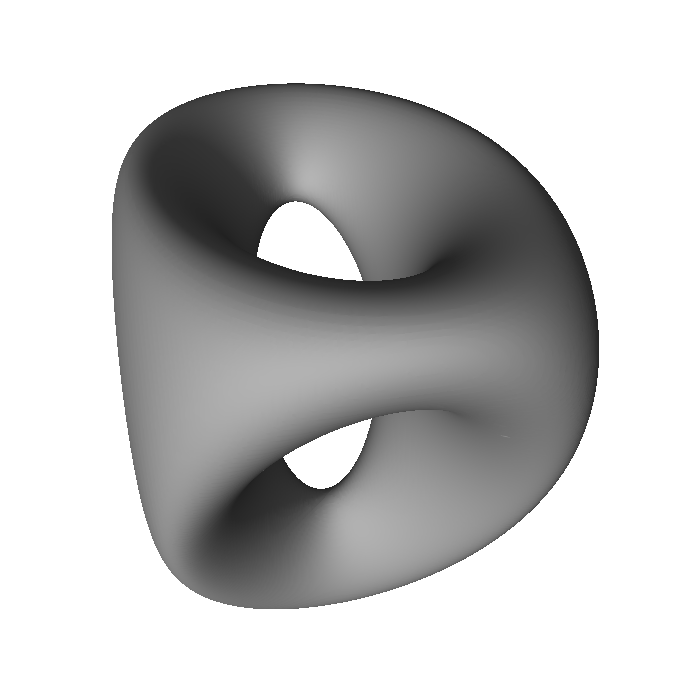}
			}
			\resizebox{0.5\textwidth}{!}
			{
				\includegraphics{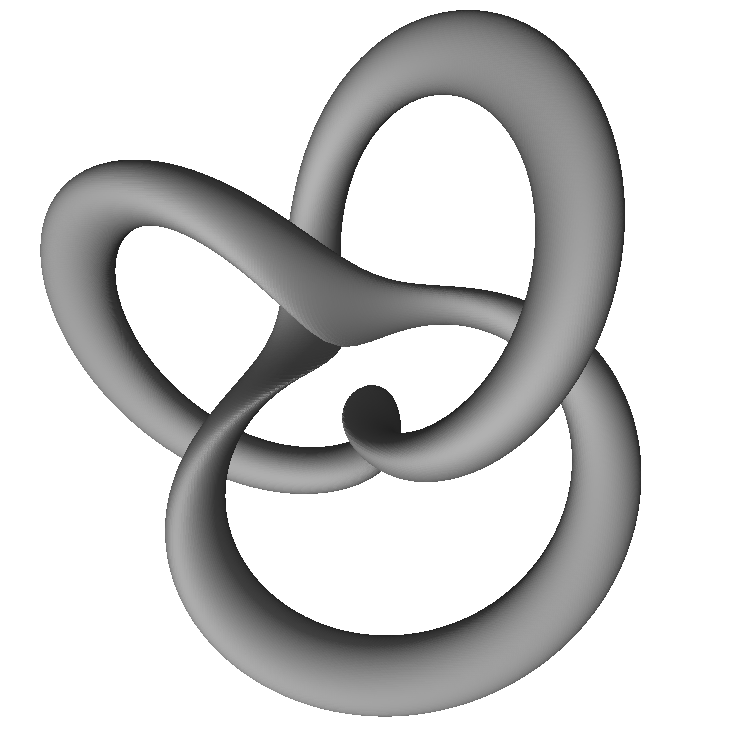}
				\includegraphics{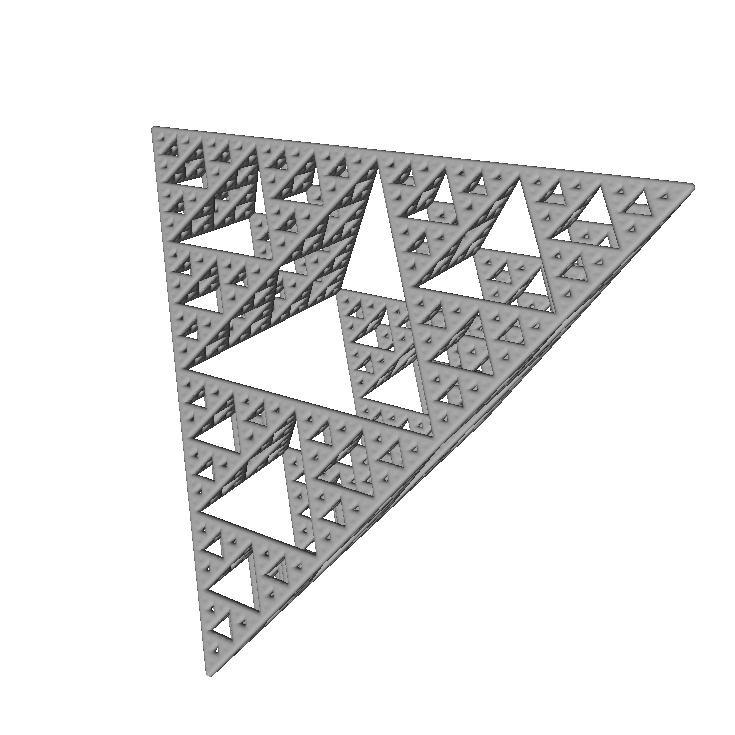}
			}
			\caption
			{
				The four scenes used in our experiments from Section \ref{sec:experiments-primitive}.
			}
			\label{fig:primitive-scenes}
		\end{figure}

		We use four different scenes in this batch of experiments.
		The first scene contains $7$ basic primitives:
		sphere, cube, cone, cylinder, torus, hexagonal prism and capsule.
		All these primitives have simple and efficient signed distance bounds.
		The second scene is a surface of genus $2$ given by the implicit equation
		$2y(y^2-3x^2)(1-z^2) + (x^2 + y^2)^2 - (9z^2 - 1)(1-z^2)=0$.
		The third scene contains a knot with an explicit signed ditance bound.
		The fourth scene contains the Sierpinski tetrahedron.
		These scenes are visualized in Figure \ref{fig:primitive-scenes}.

\begin{figure*}[ht]
			\centering
			\input{times.tikz}
			\caption
			{
				Elapsed times plotted against grid resolution for the four different scenes from Figure \ref{fig:primitive-scenes}.
				The legend for all graphs is plotted in the top left one.
				Note that all axes are \textit{logarithmic}.
			}
			\label{fig:times}
		\end{figure*}
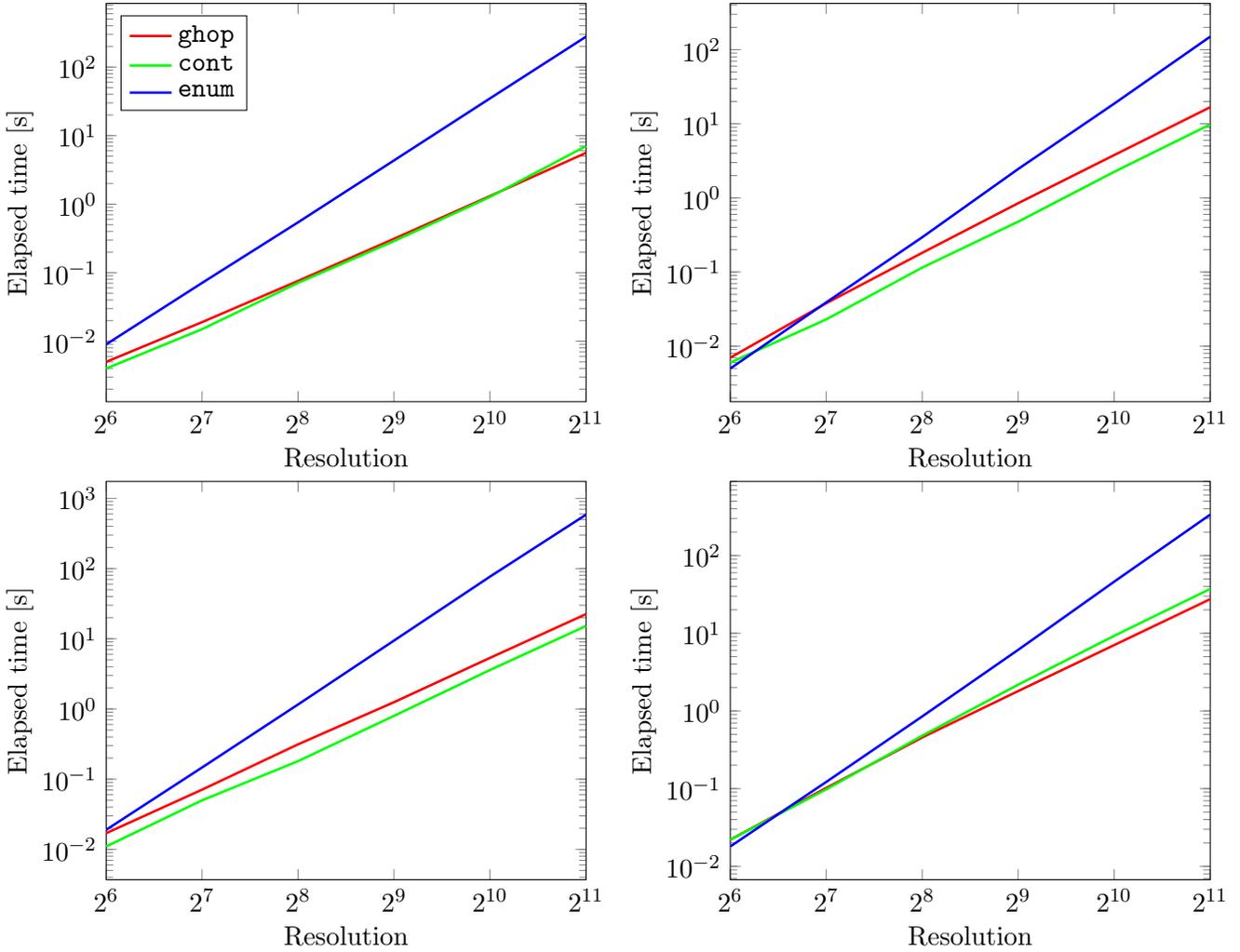

		Figure \ref{fig:times} shows the times needed to polygonize the scenes with the slow and the fast algorithm.
		Note that the axes in the graphs have logarithmic scale.
		We can see that the measured times for both methods appear as lines.
		This is expected since the computed theoretical complexities are polynomial ($\sim N^3$ and $\sim N^2$).
		However, the fast method becomes significantly faster for large $N$, i.e., asymptotically.
		This aligns with the predictions from Section \ref{sec:complexity}.

		\subsection{Experiments on learned SDBs}\label{sec:experiments-learned}
		There has been considerable interest in learning-based methods for signed distance bound modelling.
		And this is lately especially true in the area of deep learning
		\cite{GenShapeCVPR2019,DeepShapeCVPR2019,davies2020overfit,hao2020dualsdf,SecretsOfWildSDFs2021,iccvw2021,stanfordgeom2021,asdf2021,takikawa2021nglod}\footnote{See section "Implicit representation" at \url{https://github.com/subeeshvasu/Awsome_Deep_Geometry_Learning} for more (and updated) examples.}.

		Let us constrict our analysis to a set of methods that use a neural network (NN) to approximate a SDB of a shape:
		\begin{equation}\label{eq:nnsdb}
			\text{NN}_\theta(x, y, z)\approx
			d_S(x, y, z)
			,
		\end{equation}
		where $(x, y, z)\in\mathbb{R}^3$ is a point in 3D space and $d_S(x, y, z)$ denotes its Euclidean distance to the surface of the shape $S$.
		The parameters of the $NN$ are denoted with $\theta$.
		These parameters have to be tuned to make the approximation as best as possible.

		Our hypothesis is that gridhopping is a faster algorithm for transforming such neural SDBs to triangle meshes than by using the basic enumeration technique.
		This should especially be true for higher grid resolutions.

\begin{figure*}
			\centering
			\resizebox{0.8\textwidth}{!}
			{
				\includegraphics{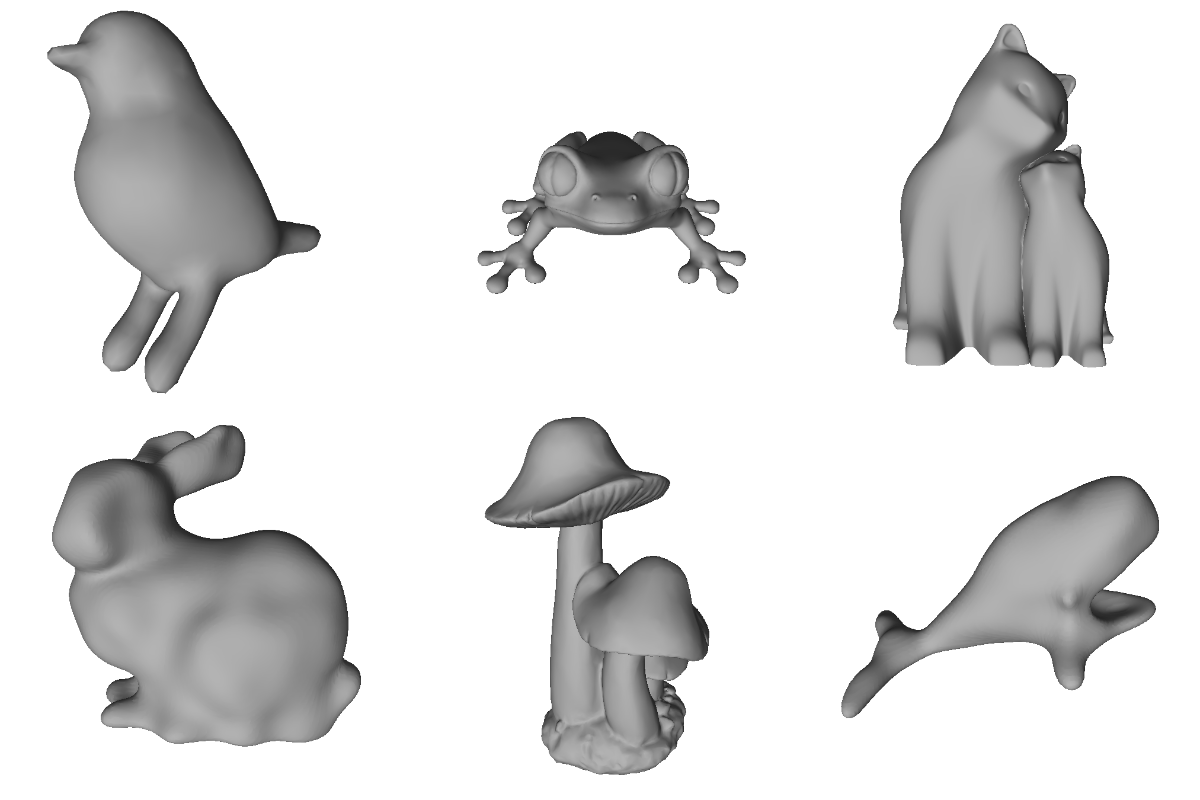}
			}
			\caption
			{
				Shapes used in experiments from Section \ref{sec:experiments-learned}.
			}
			\label{fig:3dmodels}
		\end{figure*}

\begin{figure}
			\centering
			\resizebox{0.5\textwidth}{!}
			{
				\includegraphics{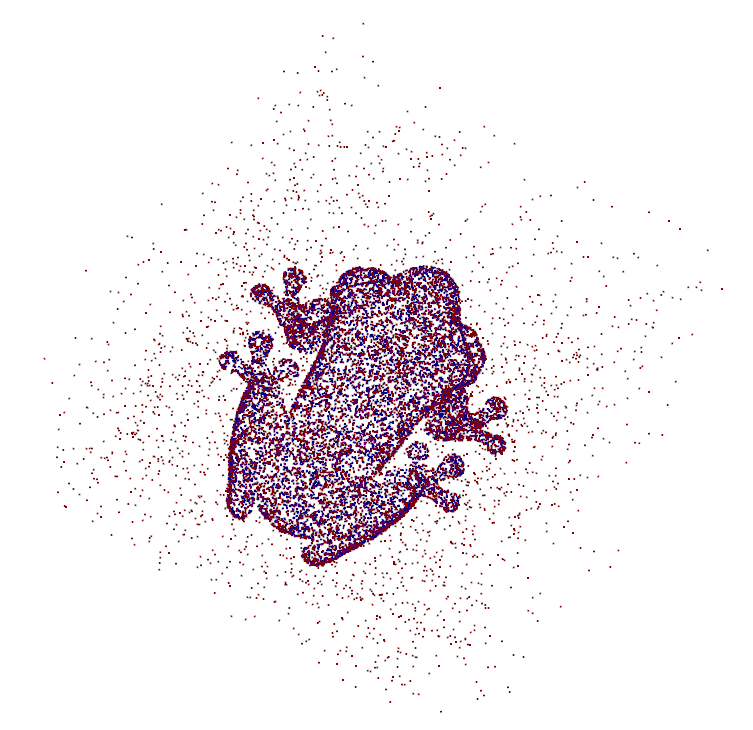}
			}
			\caption
			{
				An example point cloud generated from a 3D model of a frog.
			}
			\label{fig:ptcloud-example}
		\end{figure}

		To experimentally test our claim, we take six shapes from the Thingi10K dataset \cite{thingi10k}
		(see Figure \ref{fig:3dmodels}).
		These shapes are represented as triangle meshes and we need their SDBs.
		We first sample a dataset of point-distance pairs around each shape
		\footnote{Sampling algorithm by Park et al. \cite{DeepShapeCVPR2019}, implementation: \url{https://github.com/marian42/mesh_to_sdf}.}:
		\begin{equation}
			\{\left((x_i, y_i, z_i), d_i\right)\}_{i=1}^S
		\end{equation}
		An example can be seen in Figure \ref{fig:ptcloud-example}.
		Next, we use the Adam stochastic gradient descent technique to tune $\theta$ (from Equation \ref{eq:nnsdb}) so that $\text{NN}_\theta(x_i, y_i, z_i)\approx d_i$.
		The NN architecture is very simple: eight layers, each with embedding dimension equal to $64$, ReLU is set as the activation function.
		To speed up training, we also use batch normalization in all layers except the last one.
		This results in about $30,000$ network weights (120kB of memory).
		By modern standards, this is a tiny network, but it is capable of accurately representing the shapes used in our experiments.
		Note that our methodology trivially extends to other NN approaches for SDB modelling.
		We decided to go with this one to keep things as simple as possible.

		Recall that if certain criteria are not met (e.g., Lipschitz continuity),
		sphere tracing might "miss" parts of the surface of the shape and we get an incomplete mesh with gridhopping \cite{Hart94spheretracing}.
		In practice, this means that the NN overestiamtes the surface distance for some points
		(e.g., due to the imperfection of the learning algorithm).
		To circumvent these problems, we shrink the signed distance approximation by factor $\lambda$: $\text{SDB}(x, y, z)=\frac{\text{NN}_\theta(x, y, z)}{\lambda}$.
		This procedure slows down gridhopping and has no effects on the speed of the complete enumeration algorithm.
		%The shrinkage parameter $\lambda$ is set to $1.5$ in our experiments.
        We empirically found that $\lambda=1.5$ works well.
		This (or smaller) value results in perfect mesh reconstructions for all shapes, i.e., all methods produce identical meshes.

\begin{figure*}
			\centering
			\resizebox{1.0\textwidth}{!}
			{
			\input{nnsdb-times-cpu-1.tikz}
			}
			\resizebox{1.0\textwidth}{!}
			{
			\input{nnsdb-times-cpu-2.tikz}
			}
			\caption
			{
				CPU experiments: elapsed times plotted against grid resolution for the six different shapes from Figure \ref{fig:3dmodels}.
				The legend for all graphs is plotted in the top left one.
				Note that all axes are \textit{logarithmic}.
			}
			\label{fig:nnsdb-times-cpu}
		\end{figure*}
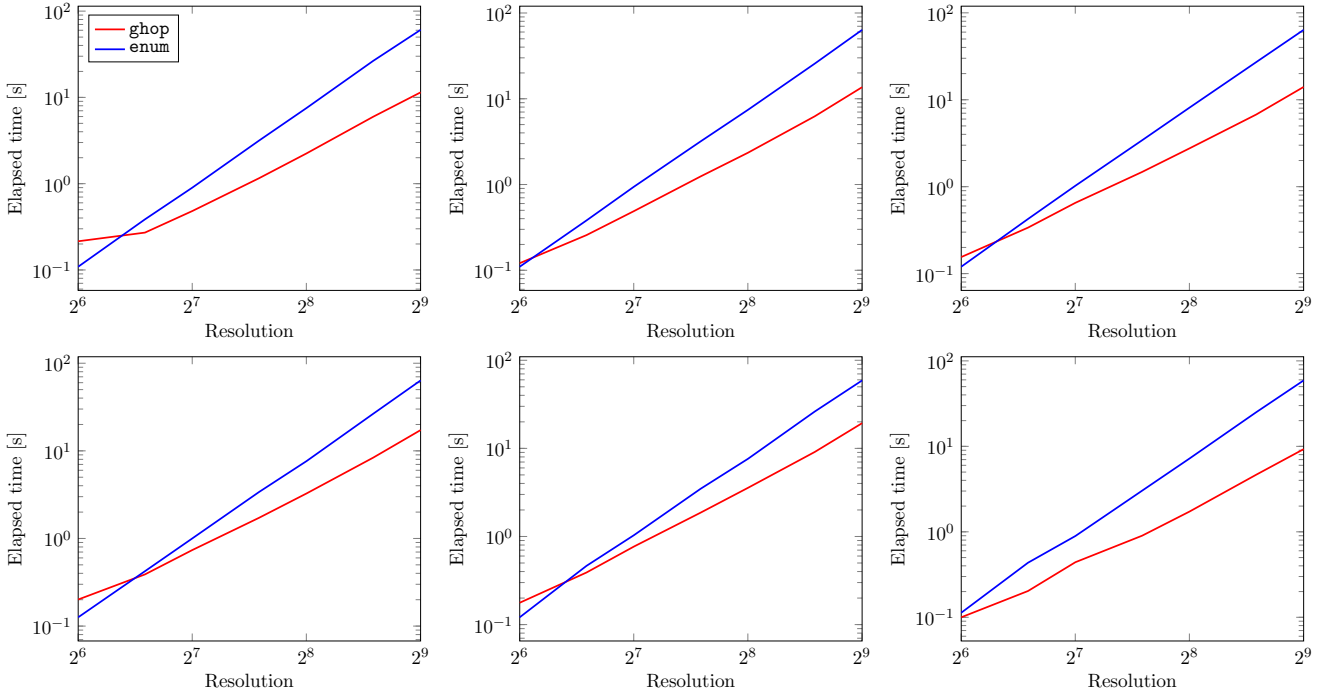

		We compare the times needed to polygonize the prepared SDBs for the complete enumeration algorithm and gridhopping.
		The grid resolution $N$ varies from $64$ to $512$.
		First, all computations are performed on Intel(R) Core(TM) i7-9750H CPU @ 2.60GHz.
		Batching the NN evaluations in chunks of size $\approx 10,000$ leads to significant speed improvements for both methods.
		The results can be seen in Figure \ref{fig:nnsdb-times-cpu}.
		Even though the enumeration method is sometimes faster for small grid resolutions, we can clearly observe that gridhopping has a significant asymptotic advantage.
		The resuts are in accordance with the theoretically predicted ones \ref{sec:complexity}: $\sim N^3$ vs. $\sim N^2$ computational complexity.

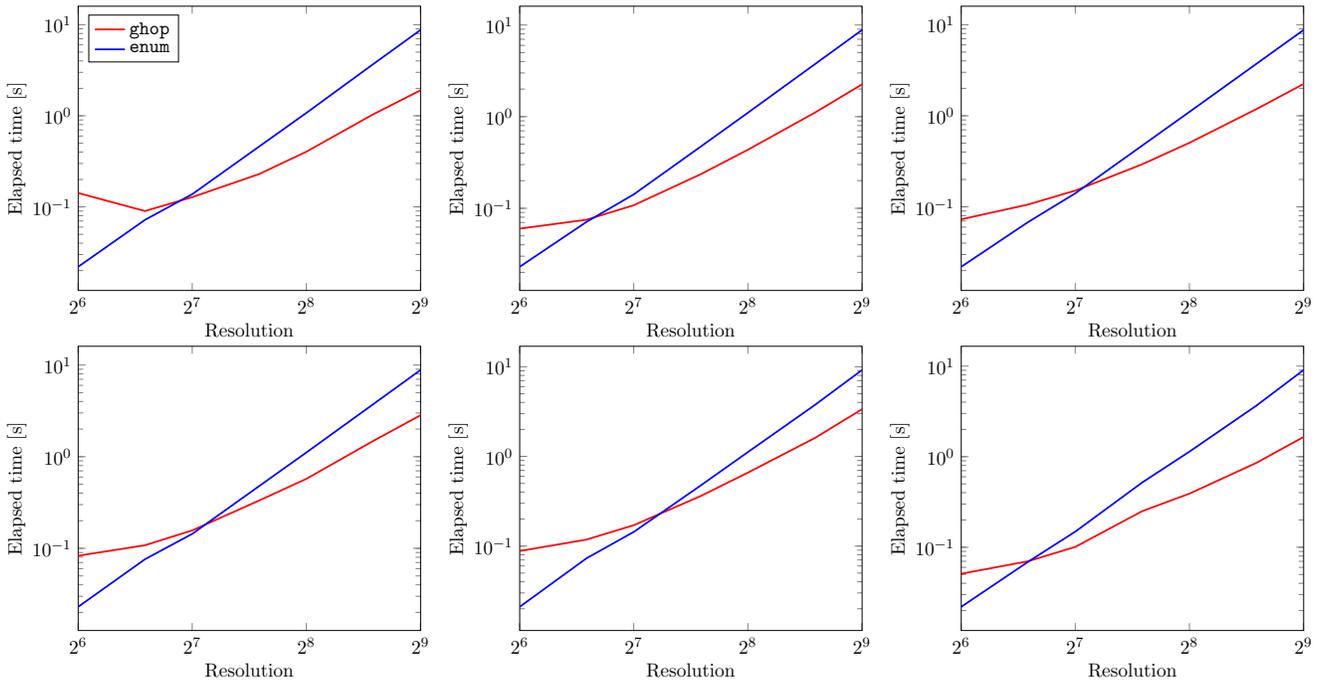
\begin{figure*}
			\centering
			\resizebox{1.0\textwidth}{!}
			{
			\input{nnsdb-times-gpu-1.tikz}
			}
			\resizebox{1.0\textwidth}{!}
			{
			\input{nnsdb-times-gpu-2.tikz}
			}
			\caption
			{
				Same setting as in Figure \ref{fig:nnsdb-times-cpu}, except the computations are done on the GPU.
			}
			\label{fig:nnsdb-times-gpu}
		\end{figure*}

		We also performed experiments when computations are performed on an Nvidia GeForce RTX 2060 Mobile GPU.
		Batching the NN computations in chunks of size $\approx 100,000$ seems best and we report timings in this setting, Figure \ref{fig:nnsdb-times-gpu}.
		The theoretical complexity results hold in this setting as well.
		Also, GPU computations are significantly faster than CPU ones.
		However, gridhopping is somewhat slower for smaller grid resolutions.
		We attribute this fact to constant overhead needed to prepare the gridhopping run.

	\section{Conclusion}\label{sec:conclusion}
		In this work we described and evaluated gridhopping -- a method for transforming shapes defined by signed distance bounds into triangle meshes.
		Its main advantages are simplicity, ease of implementation and processing speed.
        We confirmed this in theory and experiments by comparing the method to the enumeration and continuation variants of the Marching cubes algorithm \cite{LorensenCline87marchingcubes} which are the standard approaches for polygonizing signed distance bounds in practical applications.
        We used both simple shapes defined by short equations/code and more complicated ones defined through neural networks.
        We also showed that these conclusions hold when the computations are performed on the CPU and on the GPU, which is a further confirmation of the practicality of the method.

		The main limitations of gridhopping are the Lipschitz continuity requirement, that it works only with abstract, mathematically defined, shapes and not real-world data like point clouds, and the fact that it generates too many triangles for some simple shapes.
        Note that the last limitation also holds for Marching cubes as both methods produce identical triangle meshes.
        Also, within the context of this paper, we would like to mention that the Marching cubes-based pipelines (gridhopping included) might be competing with deep learning-based methods that generate triangle meshes \cite{siddiqui2023meshgpt,alliegro2023polydiff}, especially if the texture is included \cite{gao2022get3d}.
        However, these approaches are not practical in many situations due to their implementation complexity, large processing speed and the requirement for high-quality training data to be effective.
        Therefore gridhopping is still relevant for the wider engineering community.

		Our implementation of the gridhopping method is available at \url{https://github.com/nenadmarkus/gridhopping}.

    \section{Author contributions}
        Conceptualization, methodology, software, writing: N. M. Review, editing, validation, funding acquisition: M. S.  
\iffalse
        All authors have read and agreed to the published version of the manuscript.

    \section{Funding}
        This research was supported by the Croatian National Recovery and Resilience Plan (NPOO) under the project Research and Development of Multiple Innovative Products, Services and Business Models Aimed at Strengthening Sustainable Tourism and the Green and Digital Transition of Tourism (ROBOCAMP), with grant number NPOO.C1.6.R1-I2.01.
        Also, work was supported by the European Union under the project Enhanced Pilot Interfaces \& Interactions for fighter Cockpit (EPIIC) funded by European Defence Fund (EDF) Grant Agreement 101103592.

    \section{Conflicts of interest}
        The authors declare no conflicts of interest.
        The funding sponsors had no role in the design of the study; in the collection, analyses, or interpretation of data; in the writing of the manuscript, and in the decision to publish the results.
\fi

	\bibliographystyle{abbrv}
	\bibliography{references}
\end{document}

%% file: boxapprox.tikz
\begin{tikzpicture}
	% grid
	\draw[step=1.0,gray!25!,very thin] (-7, -4) grid (6, 5);

	% shape
	\draw[red, line width=1.0mm]  plot[smooth cycle, tension=.5] coordinates {(1.0, -2.0) (-1.0, 0.0) (-5.0, -2.0) (-6.0, 0.0) (2.0, 4.0) (5.0, -1.0) (4.0, -3.0) (2.0, -1.0)};

	% boxes
	\draw[blue, line width=0.25mm] (3.5, -3.0) -- (3.5, -2.5) -- (4.5, -2.5) -- (4.5, -3.0) -- cycle;
	\draw[blue, line width=0.25mm] (3.0, -2.5) -- (3.0, -2.0) -- (5.0, -2.0) -- (5.0, -2.5) -- cycle;
	\draw[blue, line width=0.25mm] (2.5, -2.0) -- (2.5, -1.5) -- (5.1, -1.5) -- (5.1, -2.0) -- cycle;
	\draw[blue, line width=0.25mm] (0.5, -2.0) -- (0.5, -1.5) -- (1.5, -1.5) -- (1.5, -2.0) -- cycle;
	\draw[blue, line width=0.25mm] (2.5, -2.0) -- (2.5, -1.5) -- (5.1, -1.5) -- (5.1, -2.0) -- cycle;
	\draw[blue, line width=0.25mm] (-6.0, -2.0) -- (-6.0, -1.5) -- (-4.0, -1.5) -- (-4.0, -2.0) -- cycle;
	\draw[blue, line width=0.25mm] (2.2, -1.5) -- (2.2, -1.0) -- (5.1, -1.0) -- (5.1, -1.5) -- cycle;
	\draw[blue, line width=0.25mm] (0.25, -1.5) -- (0.25, -1.0) -- (1.75, -1.0) -- (1.75, -1.5) -- cycle;
	\draw[blue, line width=0.25mm] (-6.25, -1.5) -- (-6.25, -1.0) -- (-3.25, -1.0) -- (-3.25, -1.5) -- cycle;
	\draw[blue, line width=0.25mm] (0.0, -1.0) -- (0.0, -0.5) -- (5.0, -0.5) -- (5.0, -1.0) -- cycle;
	\draw[blue, line width=0.25mm] (-6.4, -1.0) -- (-6.4, -0.5) -- (-2.5, -0.5) -- (-2.5, -1.0) -- cycle;
	\draw[blue, line width=0.25mm] (-6.3, -0.5) -- (-6.3, 0.0) -- (-1.8, 0.0) -- (-1.8, -0.5) -- cycle;
	\draw[blue, line width=0.25mm] (-0.5, -0.5) -- (-0.5, 0.0) -- (4.9, 0.0) -- (4.9, -0.5) -- cycle;
	\draw[blue, line width=0.25mm] (-5.9, 0.0) -- (-5.9, 0.5) -- (4.7, 0.5) -- (4.7, 0.0) -- cycle;
	\draw[blue, line width=0.25mm] (-5.2, 0.5) -- (-5.2, 1.0) -- (4.5, 1.0) -- (4.5, 0.5) -- cycle;
	\draw[blue, line width=0.25mm] (-4.2, 1.0) -- (-4.2, 1.5) -- (4.3, 1.5) -- (4.3, 1.0) -- cycle;
	\draw[blue, line width=0.25mm] (-3.5, 1.5) -- (-3.5, 2.0) -- (4.0, 2.0) -- (4.0, 1.5) -- cycle;
	\draw[blue, line width=0.25mm] (-2.5, 2.0) -- (-2.5, 2.5) -- (3.8, 2.5) -- (3.8, 2.0) -- cycle;
	\draw[blue, line width=0.25mm] (-1.5, 2.5) -- (-1.5, 3.0) -- (3.5, 3.0) -- (3.5, 2.5) -- cycle;
	\draw[blue, line width=0.25mm] (-0.5, 3.0) -- (-0.5, 3.5) -- (3.2, 3.5) -- (3.2, 3.0) -- cycle;
	\draw[blue, line width=0.25mm] (0.75, 3.5) -- (0.75, 4.0) -- (2.5, 4.0) -- (2.5, 3.5) -- cycle;

\end{tikzpicture}

%% file: times.tikz
\resizebox{0.95\textwidth}{!}
{
\begin{tikzpicture}
\begin{axis} [
	ymode=log,
	xlabel={Resolution},
	ylabel={Elapsed time [s]},
	xmin=64, xmax=2048,
    ymin=0.004, ymax=270,
	xmode=log, log basis x=2,
	legend pos=north west
]
		\addplot[color=red, line width=1]
			coordinates {
				(64.000000, 0.005000)(128.000000, 0.019000)(256.000000, 0.076000)(512.000000, 0.313000)(1024.000000, 1.314000)(2048.000000, 5.606000)
			};
		\addplot[color=green, line width=1]
			coordinates {
				(64.000000, 0.004000)(128.000000, 0.015000)(256.000000, 0.071000)(512.000000, 0.289000)(1024.000000, 1.277000)(2048.000000, 7.041000)
			};
		\addplot[color=blue, line width=1]
			coordinates {
				(64.000000, 0.009000)(128.000000, 0.071000)(256.000000, 0.544000)(512.000000, 4.350000)(1024.000000, 34.994000)(2048.000000, 277.823000)
			};
		\legend{\texttt{ghop},\texttt{cont},\texttt{enum}}
\end{axis}
\end{tikzpicture}
\begin{tikzpicture}
\begin{axis} [
	ymode=log,
	xlabel={Resolution},
	ylabel={Elapsed time [s]},
	xmin=64, xmax=2048,
    ymin=0.005, ymax=150,
	xmode=log, log basis x=2,
	legend pos=north west
]
		\addplot[color=red, line width=1]
			coordinates {
				(64.000000, 0.007000)(128.000000, 0.038000)(256.000000, 0.182000)(512.000000, 0.850000)(1024.000000, 3.786000)(2048.000000, 16.752000)
			};
		\addplot[color=green, line width=1]
			coordinates {
				(64.000000, 0.006000)(128.000000, 0.023000)(256.000000, 0.115000)(512.000000, 0.479000)(1024.000000, 2.247000)(2048.000000, 9.802000)
			};
		\addplot[color=blue, line width=1]
			coordinates {
				(64.000000, 0.005000)(128.000000, 0.039000)(256.000000, 0.295000)(512.000000, 2.459000)(1024.000000, 18.711000)(2048.000000, 149.853000)
			};
\end{axis}
\end{tikzpicture}
}

\resizebox{0.95\textwidth}{!}
{
\begin{tikzpicture}
\begin{axis} [
	ymode=log,
	xlabel={Resolution},
	ylabel={Elapsed time [s]},
	xmin=64, xmax=2048,
    ymin=0.011, ymax=400,
	xmode=log, log basis x=2,
	legend pos=north west
]
		\addplot[color=red, line width=1]
			coordinates {
				(64.000000, 0.017000)(128.000000, 0.071000)(256.000000, 0.314000)(512.000000, 1.252000)(1024.000000, 5.351000)(2048.000000, 22.459000)
			};
		\addplot[color=green, line width=1]
			coordinates {
				(64.000000, 0.011000)(128.000000, 0.05000)(256.000000, 0.181000)(512.000000, 0.800000)(1024.000000, 3.594000)(2048.000000, 15.191000)
			};
		\addplot[color=blue, line width=1]
			coordinates {
				(64.000000, 0.019000)(128.000000, 0.147000)(256.000000, 1.161000)(512.000000, 9.435000)(1024.000000, 76.977000)(2048.000000, 586.260000)
			};
\end{axis}
\end{tikzpicture}
\begin{tikzpicture}
\begin{axis} [
	ymode=log,
	xlabel={Resolution},
	ylabel={Elapsed time [s]},
	xmin=64, xmax=2048,
    ymin=0.018, ymax=300,
	xmode=log, log basis x=2,
	legend pos=north west
]
		\addplot[color=red, line width=1]
			coordinates {
				(64.000000, 0.022000)(128.000000, 0.102000)(256.000000, 0.456000)(512.000000, 1.805000)(1024.000000, 7.040000)(2048.000000, 27.455000)
			};
		\addplot[color=green, line width=1]
			coordinates {
				(64.000000, 0.022000)(128.000000, 0.098000)(256.000000, 0.480000)(512.000000, 2.182000)(1024.000000, 9.262000)(2048.000000, 37.213000)
			};
		\addplot[color=blue, line width=1]
			coordinates {
				(64.000000, 0.018000)(128.000000, 0.122000)(256.000000, 0.851000)(512.000000, 6.121000)(1024.000000, 46.149000)(2048.000000, 336.216000)
			};
\end{axis}
\end{tikzpicture}
}

%% file: nnsdb-times-cpu-1.tikz
\begin{tikzpicture}
\begin{axis} [
	ymode=log,
	xlabel={Resolution},
	ylabel={Elapsed time [s]},
	xmin=64, xmax=512,
	xmode=log, log basis x=2,
	legend pos=north west
]
		\addplot[color=red, line width=1]
			coordinates {
				(64.000000, 0.215000)(96.000000, 0.271000)(128.000000, 0.482000)(192.000000, 1.157000)(256.000000, 2.241000)(384.000000, 5.969000)(512.000000, 11.437000)
			};
		\addplot[color=blue, line width=1]
			coordinates {
				(64.000000, 0.109000)(96.000000, 0.386000)(128.000000, 0.900000)(192.000000, 3.160000)(256.000000, 7.523000)(384.000000, 26.518000)(512.000000, 60.646000)
			};
		\legend{\texttt{ghop},\texttt{enum}}
\end{axis}
\end{tikzpicture}
\begin{tikzpicture}
\begin{axis} [
	ymode=log,
	xlabel={Resolution},
	ylabel={Elapsed time [s]},
	xmin=64, xmax=512,
	xmode=log, log basis x=2,
	legend pos=north west
]
		\addplot[color=red, line width=1]
			coordinates {
				(64.000000, 0.121000)(96.000000, 0.257000)(128.000000, 0.489000)(192.000000, 1.242000)(256.000000, 2.348000)(384.000000, 6.241000)(512.000000, 13.644000)
			};
		\addplot[color=blue, line width=1]
			coordinates {
				(64.000000, 0.110000)(96.000000, 0.381000)(128.000000, 0.941000)(192.000000, 3.181000)(256.000000, 7.432000)(384.000000, 25.675000)(512.000000, 63.580000)
			};
\end{axis}
\end{tikzpicture}
\begin{tikzpicture}
\begin{axis} [
	ymode=log,
	xlabel={Resolution},
	ylabel={Elapsed time [s]},
	xmin=64, xmax=512,
	xmode=log, log basis x=2,
	legend pos=north west
]
		\addplot[color=red, line width=1]
			coordinates {
				(64.000000, 0.155000)(96.000000, 0.338000)(128.000000, 0.652000)(192.000000, 1.477000)(256.000000, 2.759000)(384.000000, 6.764000)(512.000000, 14.040000)
			};
		\addplot[color=blue, line width=1]
			coordinates {
				(64.000000, 0.120000)(96.000000, 0.425000)(128.000000, 1.025000)(192.000000, 3.420000)(256.000000, 8.117000)(384.000000, 27.117000)(512.000000, 63.694000)
			};
\end{axis}
\end{tikzpicture}

%% file: nnsdb-times-cpu-2.tikz
\begin{tikzpicture}
\begin{axis} [
	ymode=log,
	xlabel={Resolution},
	ylabel={Elapsed time [s]},
	xmin=64, xmax=512,
	xmode=log, log basis x=2,
	legend pos=north west
]
		\addplot[color=red, line width=1]
			coordinates {
				(64.000000, 0.201000)(96.000000, 0.389000)(128.000000, 0.738000)(192.000000, 1.722000)(256.000000, 3.261000)(384.000000, 8.366000)(512.000000, 17.258000)
			};
		\addplot[color=blue, line width=1]
			coordinates {
				(64.000000, 0.126000)(96.000000, 0.423000)(128.000000, 1.002000)(192.000000, 3.393000)(256.000000, 7.629000)(384.000000, 26.501000)(512.000000, 63.645000)
			};
\end{axis}
\end{tikzpicture}
\begin{tikzpicture}
\begin{axis} [
	ymode=log,
	xlabel={Resolution},
	ylabel={Elapsed time [s]},
	xmin=64, xmax=512,
	xmode=log, log basis x=2,
	legend pos=north west
]
		\addplot[color=red, line width=1]
			coordinates {
				(64.000000, 0.177000)(96.000000, 0.390000)(128.000000, 0.768000)(192.000000, 1.864000)(256.000000, 3.567000)(384.000000, 9.102000)(512.000000, 19.336000)
			};
		\addplot[color=blue, line width=1]
			coordinates {
				(64.000000, 0.121000)(96.000000, 0.464000)(128.000000, 1.033000)(192.000000, 3.478000)(256.000000, 7.617000)(384.000000, 26.181000)(512.000000, 59.055000)
			};
\end{axis}
\end{tikzpicture}
\begin{tikzpicture}
\begin{axis} [
	ymode=log,
	xlabel={Resolution},
	ylabel={Elapsed time [s]},
	xmin=64, xmax=512,
	xmode=log, log basis x=2,
	legend pos=north west
]
		\addplot[color=red, line width=1]
			coordinates {
				(64.000000, 0.100000)(96.000000, 0.203000)(128.000000, 0.441000)(192.000000, 0.901000)(256.000000, 1.726000)(384.000000, 4.660000)(512.000000, 9.272000)
			};
		\addplot[color=blue, line width=1]
			coordinates {
				(64.000000, 0.113000)(96.000000, 0.435000)(128.000000, 0.893000)(192.000000, 3.030000)(256.000000, 7.206000)(384.000000, 25.092000)(512.000000, 59.058000)
			};
\end{axis}
\end{tikzpicture}

%% file: nnsdb-times-gpu-1.tikz
\begin{tikzpicture}
\begin{axis} [
	ymode=log,
	xlabel={Resolution},
	ylabel={Elapsed time [s]},
	xmin=64, xmax=512,
	xmode=log, log basis x=2,
	legend pos=north west
]
		\addplot[color=red, line width=1]
			coordinates {
				(64.000000, 0.142000)(96.000000, 0.090000)(128.000000, 0.128000)(192.000000, 0.229000)(256.000000, 0.404000)(384.000000, 1.038000)(512.000000, 1.905000)
			};
		\addplot[color=blue, line width=1]
			coordinates {
				(64.000000, 0.022000)(96.000000, 0.072000)(128.000000, 0.138000)(192.000000, 0.459000)(256.000000, 1.077000)(384.000000, 3.693000)(512.000000, 8.765000)
			};
		\legend{\texttt{ghop},\texttt{enum}}
\end{axis}
\end{tikzpicture}
\begin{tikzpicture}
\begin{axis} [
	ymode=log,
	xlabel={Resolution},
	ylabel={Elapsed time [s]},
	xmin=64, xmax=512,
	xmode=log, log basis x=2,
	legend pos=north west
]
		\addplot[color=red, line width=1]
			coordinates {
				(64.000000, 0.060000)(96.000000, 0.075000)(128.000000, 0.108000)(192.000000, 0.235000)(256.000000, 0.436000)(384.000000, 1.103000)(512.000000, 2.253000)
			};
		\addplot[color=blue, line width=1]
			coordinates {
				(64.000000, 0.023000)(96.000000, 0.071000)(128.000000, 0.142000)(192.000000, 0.470000)(256.000000, 1.105000)(384.000000, 3.726000)(512.000000, 8.842000)
			};
\end{axis}
\end{tikzpicture}
\begin{tikzpicture}
\begin{axis} [
	ymode=log,
	xlabel={Resolution},
	ylabel={Elapsed time [s]},
	xmin=64, xmax=512,
	xmode=log, log basis x=2,
	legend pos=north west
]
		\addplot[color=red, line width=1]
			coordinates {
				(64.000000, 0.073000)(96.000000, 0.106000)(128.000000, 0.151000)(192.000000, 0.293000)(256.000000, 0.505000)(384.000000, 1.181000)(512.000000, 2.235000)
			};
		\addplot[color=blue, line width=1]
			coordinates {
				(64.000000, 0.022000)(96.000000, 0.068000)(128.000000, 0.141000)(192.000000, 0.470000)(256.000000, 1.104000)(384.000000, 3.706000)(512.000000, 8.756000)
			};
\end{axis}
\end{tikzpicture}

%% file: nnsdb-times-gpu-2.tikz
\begin{tikzpicture}
\begin{axis} [
	ymode=log,
	xlabel={Resolution},
	ylabel={Elapsed time [s]},
	xmin=64, xmax=512,
	xmode=log, log basis x=2,
	legend pos=north west
]
		\addplot[color=red, line width=1]
			coordinates {
				(64.000000, 0.083000)(96.000000, 0.108000)(128.000000, 0.157000)(192.000000, 0.332000)(256.000000, 0.574000)(384.000000, 1.484000)(512.000000, 2.836000)
			};
		\addplot[color=blue, line width=1]
			coordinates {
				(64.000000, 0.023000)(96.000000, 0.076000)(128.000000, 0.144000)(192.000000, 0.477000)(256.000000, 1.113000)(384.000000, 3.738000)(512.000000, 8.840000)
			};
\end{axis}
\end{tikzpicture}
\begin{tikzpicture}
\begin{axis} [
	ymode=log,
	xlabel={Resolution},
	ylabel={Elapsed time [s]},
	xmin=64, xmax=512,
	xmode=log, log basis x=2,
	legend pos=north west
]
		\addplot[color=red, line width=1]
			coordinates {
				(64.000000, 0.088000)(96.000000, 0.118000)(128.000000, 0.171000)(192.000000, 0.361000)(256.000000, 0.660000)(384.000000, 1.601000)(512.000000, 3.361000)
			};
		\addplot[color=blue, line width=1]
			coordinates {
				(64.000000, 0.021000)(96.000000, 0.073000)(128.000000, 0.144000)(192.000000, 0.473000)(256.000000, 1.119000)(384.000000, 3.737000)(512.000000, 9.211000)
			};
\end{axis}
\end{tikzpicture}
\begin{tikzpicture}
\begin{axis} [
	ymode=log,
	xlabel={Resolution},
	ylabel={Elapsed time [s]},
	xmin=64, xmax=512,
	xmode=log, log basis x=2,
	legend pos=north west
]
		\addplot[color=red, line width=1]
			coordinates {
				(64.000000, 0.051000)(96.000000, 0.070000)(128.000000, 0.101000)(192.000000, 0.250000)(256.000000, 0.391000)(384.000000, 0.853000)(512.000000, 1.652000)
			};
		\addplot[color=blue, line width=1]
			coordinates {
				(64.000000, 0.022000)(96.000000, 0.069000)(128.000000, 0.149000)(192.000000, 0.518000)(256.000000, 1.138000)(384.000000, 3.649000)(512.000000, 9.065000)
			};
\end{axis}
\end{tikzpicture}

%% file: body.bbl
\begin{thebibliography}{10}

\bibitem{alliegro2023polydiff}
A.~Alliegro, Y.~Siddiqui, T.~Tommasi, and M.~Nießner.
\newblock Polydiff: Generating 3d polygonal meshes with diffusion models, 2023.

\bibitem{bloomenthal}
J.~Bloomenthal, editor.
\newblock {\em Introduction to Implicit Surfaces}.
\newblock Morgan-Kaufmann, 1997.

\bibitem{GenShapeCVPR2019}
Z.~Chen and H.~Zhang.
\newblock Learning implicit fields for generative shape modeling.
\newblock In {\em CVPR}, 2019.

\bibitem{CigoniEtAl}
P.~Cigoni, P.~Marino, C.~Montani, E.~Puppo, and R.~Scopigno.
\newblock Speeding up isosurface extraction using interval trees.
\newblock {\em IEEE Transactions on Visualization and Computer Graphics}, 1997.

\bibitem{davies2020overfit}
T.~Davies, D.~Nowrouzezahrai, and A.~Jacobson.
\newblock Overfit neural networks as a compact shape representation.
\newblock \url{https://arxiv.org/abs/2009.09808}, 2020.

\bibitem{SecretsOfWildSDFs2021}
S.~Duggal, Z.~Wang, W.-C. Ma, S.~Manivasagam, J.~Liang, S.~Wang, and
  R.~Urtasun.
\newblock Secrets of 3d implicit object shape reconstruction in the wild.
\newblock \url{https://arxiv.org/abs/2101.06860}, 2021.

\bibitem{gao2022get3d}
J.~Gao, T.~Shen, Z.~Wang, W.~Chen, K.~Yin, D.~Li, O.~Litany, Z.~Gojcic, and
  S.~Fidler.
\newblock Get3d: A generative model of high quality 3d textured shapes learned
  from images.
\newblock In {\em Advances In Neural Information Processing Systems}, 2022.

\bibitem{hao2020dualsdf}
Z.~Hao, H.~Averbuch-Elor, N.~Snavely, and S.~Belongie.
\newblock Dualsdf: Semantic shape manipulation using a two-level
  representation.
\newblock In {\em CVPR}, 2020.

\bibitem{Hart94spheretracing}
J.~C. Hart.
\newblock Sphere tracing: A geometric method for the antialiased ray tracing of
  implicit surfaces.
\newblock {\em The Visual Computer}, 1994.

\bibitem{HartEtAl89rtfractals}
J.~C. Hart, D.~J. Sandin, and L.~H. Kauffman.
\newblock Ray tracing deterministic 3-d fractals.
\newblock {\em ACM SIGGRAPH computer graphics}, 1989.

\bibitem{ho2020denoising}
J.~Ho, A.~Jain, and P.~Abbeel.
\newblock Denoising diffusion probabilistic models, 2020.

\bibitem{ShapeFromPtCould}
L.~Li, M.~Sung, A.~Dubrovina, L.~Yi, and L.~Guibas.
\newblock Supervised fitting of geometric primitives to 3d point clouds.
\newblock In {\em CVPR}, 2019.

\bibitem{LivnatShenJohnson}
Y.~Livnat, H.-W. Shen, and C.~R. Johnson.
\newblock A near optimal isosurface extraction algorithm using the span space.
\newblock {\em IEEE Transactions on Visualization and Computer Graphics}, 1996.

\bibitem{LorensenCline87marchingcubes}
W.~E. Lorensen and H.~E. Cline.
\newblock Marching cubes: A high resolution 3d surface construction algorithm.
\newblock {\em ACM SIGGRAPH computer graphics}, 1987.

\bibitem{asdf2021}
J.~Mu, W.~Qiu, A.~Kortylewski, A.~Yuille, N.~Vasconcelos, and X.~Wang.
\newblock A-sdf: Learning disentangled signed distance functions for
  articulated shape representation.
\newblock \url{https://arxiv.org/abs/2104.07645}, 2021.

\bibitem{olahImplicitCAD}
C.~Olah.
\newblock Implicitcad.
\newblock \url{https://github.com/Haskell-Things/ImplicitCAD}, 2011.
\newblock Accessed: Feb 26 2024.

\bibitem{olah2011manipulation}
C.~Olah.
\newblock Manipulation of implicit functions with an eye on cad.
\newblock
  \url{https://christopherolah.wordpress.com/2011/11/06/manipulation-of-implicit-functions-with-an-eye-on-cad/},
  2011.
\newblock Accessed: Feb 26 2024.

\bibitem{DeepShapeCVPR2019}
J.~J. Park, P.~Florence, J.~Straub, R.~Newcombe, and S.~Lovegrove.
\newblock Deepsdf: Learning continuous signed distance functionsfor shape
  representation.
\newblock In {\em CVPR}, 2019.

\bibitem{siddiqui2023meshgpt}
Y.~Siddiqui, A.~Alliegro, A.~Artemov, T.~Tommasi, D.~Sirigatti, V.~Rosov,
  A.~Dai, and M.~Nießner.
\newblock Meshgpt: Generating triangle meshes with decoder-only transformers,
  2023.

\bibitem{sohldickstein2015deep}
J.~Sohl-Dickstein, E.~Weiss, N.~Maheswaranathan, and S.~Ganguli.
\newblock Deep unsupervised learning using nonequilibrium thermodynamics.
\newblock In {\em ICML}, 2015.

\bibitem{takikawa2021nglod}
T.~Takikawa, J.~Litalien, K.~Yin, K.~Kreis, C.~Loop, D.~Nowrouzezahrai,
  A.~Jacobson, M.~McGuire, and S.~Fidler.
\newblock Neural geometric level of detail: Real-time rendering with implicit
  {3D} shapes.
\newblock In {\em CVPR}, 2021.

\bibitem{hseries}
E.~W. Weisstein.
\newblock Harmonic series. {From MathWorld---A Wolfram Web Resource}.
\newblock Accessed: Feb 26 2024.

\bibitem{ppdist}
E.~W. Weisstein.
\newblock Point-plane distance. {From MathWorld---A Wolfram Web Resource}.
\newblock Accessed: Feb 26 2024.

\bibitem{WilhelmsGelder}
J.~Wilhelms and A.~V. Gelder.
\newblock Octrees for faster isosurface generation.
\newblock {\em Transactions on Graphics}, 1992.

\bibitem{DoubleLayerSDF}
Y.~Wu, J.~Man, and Z.~Xie.
\newblock A double layer method for constructing signed distance fields from
  triangle meshes.
\newblock {\em Graphical Models}, 2014.

\bibitem{iccvw2021}
S.~Yao, F.~Yang, Y.~Cheng, and M.~G. Mozerov.
\newblock 3d shapes local geometry codes learning with sdf.
\newblock In {\em Proceedings of the ICCV Workshops}, 2021.

\bibitem{stanfordgeom2021}
W.~Yifan, L.~Rahmann, and O.~Sorkine-Hornung.
\newblock Geometry-consistent neural shape representation with implicit
  displacement fields.
\newblock \url{https://arxiv.org/abs/2106.05187}, 2021.

\bibitem{thingi10k}
Q.~Zhou and A.~Jacobson.
\newblock Thingi10k: A dataset of 10,000 3d-printing models.
\newblock \url{https://arxiv.org/abs/1605.04797}, 2016.

\end{thebibliography}
